\documentstyle[aps,pre,preprint,epsf]{revtex}

\newcommand{\eqref}[1]{Eq.~(\protect\ref{#1})}

\begin{document}

\title{Failure of linear control in noisy coupled map lattices}
\author{David~A. Egolf} 
\address{Laboratory of Atomic and Solid State Physics,
         Cornell University, Ithaca, NY 14853}
\author{Joshua E. S. Socolar} 
\address{Department of Physics and Center for Nonlinear and Complex Systems,
         Duke University, Durham, NC 27708}
\date{December 18, 1997}

\maketitle

\begin{abstract}

We study a 1D ring of diffusively coupled logistic maps in the
vicinity of an unstable, spatially homogeneous fixed point.  The
failure of linear controllers due to additive noise is discussed with
the aim of clarifying the failure mechanism.  A criterion is suggested
for estimating the noise level that can be tolerated by the given
controller.  The criterion implies the loss of control for
surprisingly low noise levels in certain cases of interest, and
accurately accounts for the results of numerical experiments over a
broad range of parameter values.  Previous results of Grigoriev {\it
et al.} (Phys. Rev. Lett. {\bf 79}, 2795) are reviewed and compared
with our numerical and analytic results.

\end{abstract}


\narrowtext

\section{Introduction}

Over the past several years, work on the feedback stabilization of
periodic orbits in nonlinear dynamical systems has heightened interest
within the physics community in the characteristics of various types
of feedback-controlled systems.  A problem of particular importance to
physicists is the stabilization of uniform or ordered states in
spatially extended systems, and it is of interest to analyze generic
models of such systems and elucidate the features that make control
difficult.  Two recent articles have examined what is arguably the
simplest representation of a spatially extended system with an
unstable, homogeneous fixed point --- a 1D ring of diffusively coupled
logistic maps \cite{huprl,grigorievprl}. This paper extends those
analyses with the aim of clarifying the mechanism responsible for the
failure of linear control in the presence of additive noise.  Though
more rigorous mathematical techniques of the theory of robust control
can be applied to this problem to address specific engineering
objectives \cite{doyle}, the following analysis provides a useful
conceptual picture of the behavior that can be expected to be
quantitatively accurate for generic physical systems.

Following Grigoriev {\it et al.} \cite{grigorievprl}, 
we consider the feedback control of a ring of 
diffusively coupled maps with additive white noise:
\begin{equation}\label{efull}
z_i^{(t+1)} = f\left(z_i^{(t)} + 
        \epsilon (z_{i-1}^{(t)} - 2 z_i^{(t)} + z_{i+1}^{(t)})\right)
              + \eta_i^{(t)},
\end{equation}
where subscripts indicate spatial position and 
superscripts in parentheses indicate temporal iterates.
(Throughout this paper superscripts {\em without} parentheses will
indicate exponents.)
For a given ring size, we show how a typical controller designed using
standard linear-quadratic control theory may fail in the presence of 
very low noise levels, and we suggest a criterion for estimating 
the maximum tolerable noise level.
For a fixed noise level, the maximum ring size that can be controlled
can be taken as an estimate of the maximum 
allowable spacing between controllers
in a much larger ring.

The noise $\eta_i^{(t)}$ is taken to be an independent number
between $-\sqrt{3} \sigma$
and $\sqrt{3} \sigma$, so we have
\begin{eqnarray} 
\left\langle\eta_i^{(t)}\right\rangle & = & 
        0\quad\forall i,t \label{eeta1}\\
\left\langle\eta_i^{(t)}\eta_j^{(s)}\right\rangle & = & 
        \sigma^2\delta_{ij}\delta_{ts},
 \label{eeta2}
\end{eqnarray}
where $\langle\rangle$ represents an ensemble average.
We take the ring to contain $L$ sites and the spatial index to run
from $1$ to $L$, so
$z_0$ is identified with $z_L$, and $z_{L+1}$ with $z_1$. 
For concreteness, we will take $f$ to be the logistic map
\begin{equation}
f(z) = \mu z (1-z),
\end{equation}
where $\mu$ is a real parameter.
Generalization of the results to other maps is straightforward.

For any positive $\mu$, \eqref{efull} has a homogeneous fixed point
solution 
$z_i = 1 - 1/\mu \equiv z^\ast$.  
For $\mu>3$, the homogeneous solution continues to exist
but is unstable to long wavelength fluctuations.
To study the stability of this solution,
we linearize the system in the vicinity of the fixed point.
Letting $x_i^{(t)} = z_i^{(t)} - z^\ast$, we obtain
\begin{equation}\label{elinear}
{\bf x}^{(t+1)} = {\bf A}\cdot{\bf x}^{(t)},
\end{equation}
with 
\begin{equation}
{\bf A} = \alpha\left(\begin{array}{cccccc}
 1-2\epsilon & \epsilon    & 0        & \cdots   & 0           & \epsilon \\
 \epsilon    & 1-2\epsilon & \epsilon & \ddots   & \ddots      & 0 \\
 0           & \ddots      & \ddots   & \ddots   & \ddots      & \vdots \\
 \vdots      & \ddots      & \ddots   & \ddots   & \ddots      & 0 \\
 0           & \ddots      & \ddots   & \epsilon & 1-2\epsilon & \epsilon \\
 \epsilon    & 0           & \cdots   & 0        & \epsilon    & 1-2\epsilon
                      \end{array}\right),
\end{equation}
where $\alpha = 2 - \mu$ is the Floquet multiplier of $f$ at the fixed point.

Grigoriev {\it et al.} have applied well-known methods to show that
control can be achieved in this noiseless system for arbitrarily large
$L$ with just two controllers placed at adjacent sites
\cite{grigorievprl}.  Taking the two sites to be $i=1$ and $i=L$, the
controlled system in the linear regime is written as 
\begin{equation}
\label{eABK}
{\bf x}^{(t+1)} = ({\bf A} - {\bf B K}){\bf \cdot x}^{(t)},
\end{equation}
where ${\bf B}$ is a $2\times L$ matrix with
$B_{11}=B_{2L}=1$ and all other elements 0, and ${\bf K}$ is a
$L\times 2$ matrix determined by iterative solution of an appropriate
Ricatti equation derived using standard techniques of linear-quadratic
control theory \cite{ogata1,ogata2}.  We emphasize that the control
scheme requires that every site of the system be observed; the
feedback signals at sites 1 and $L$ are formed from a linear
combination of all the $x_i$'s on the most recent time step.  The
problem of determining the optimal configuration for local controllers
that only receive local information is beyond the scope of this work.

Grigoriev {\it et al.} have also pointed out that arbitrarily low
noise levels destroy the control for sufficiently large system sizes.
In general, for any given noise level $\sigma$, the full nonlinear
system will not be stabilized by the feedback of \eqref{eABK} for
sufficiently large $L$, or for fixed $L$ and sufficiently large
$\sigma$.  As will be discussed below, the breakdown of control is
caused solely by the fact that the linear-quadratic control theory
based on \eqref{elinear} does not take account of the nonlinear terms
in \eqref{efull}.  Naively, one might expect that the nonlinear
deviations would be of order $\sigma^2$ and hence never play an
important role for small $\sigma$, but it turns out that the feedback
matrix ${\bf B K}$ necessarily becomes increasingly singular with
increasing $L$, leading to great amplification of the noise by the
controller itself.  The nonlinear deviations due to this amplified
noise can be large compared to the original noise level.  When this
occurs, the nonlinear deviations themselves are amplified further by
the controller and the system quickly ``blows up''.

In the following, we first describe a method for calculating the 
amplification of the noise by the controller.
We then present our explicit criterion for estimating the tolerable noise
level and show that it compares well with the numerical data.
Finally, we compare our estimate to a different one suggested by 
Grigoriev {\it et al.},
pointing out the relative advantages and disadvantages of each.

\section{Noise Amplification}

The amplification of noise by the controller is a purely linear effect
due to the nonnormality of the eigenvectors of the matrix ${\bf A-B
K}$.  In a nonnormal system that is linearly stable about ${\bf x}=0$
small perturbations in some directions may lead to transient growth in
$||{\bf x}||$ before the eventual exponential decay.  If the degree of
nonnormality is high and the relevant eigenvalues are not too close to
being degenerate, one can obtain large transient amplifications of an
initial perturbation.  Trefethen has emphasized the destabilizing
influence of nonlinearities on highly nonnormal systems
\cite{trefethen1,trefethen3}.

Nonnormality is intrinsic to the problem of
controlling a spatially extended system using sparsely distributed
actuators.
It is intuitively obvious that a perturbation occuring far from
any actuator will undergo transient growth before the feedback
generated by the controller can propagate to the position where
the control is needed.
This transient growth in the controlled system
can be accounted for only by the
occurence of nonnormal eigenvectors in the problem.
As this physical picture suggests, the nonnormal effects become
increasingly important with increasing $L$ and a fixed number of
controllers.

As discussed by Trefethen and others, a nonnormal matrix can
be characterized by its $\epsilon$-pseudospectra, which can be
used to place bounds on the size of the transient growth of
initial deviations from the fixed point.
We are not aware, however, of any analytic techniques for
determining these pseudospectra and so turn instead to
a straightforward calculation of the noise amplification,
taking advantage of the assumed delta-function correlations
in the noise.

For the system defined by \eqref{eABK}, we
define an amplification constant $\gamma$ by the equation
\begin{equation} \label{egammadef}
\lim_{T\rightarrow\infty}
\left\langle\frac{1}{L}|{\bf x}^{(T)}|^2\right\rangle^{1/2} = \gamma\sigma.
\end{equation}
Given an explicit form for the matrix ${\bf A}-{\bf B K}$,
$\gamma$ can be computed as follows.

Let ${\bf M} \equiv {\bf A}-{\bf B K}$, let ${\bf e}_{(i)}$ be
a normalized eigenvector of ${\bf M}$, where $1<i<L$,
let ${\bf v}_{(i)}$ be
the vector orthogonal to all of the ${\bf e}_{(j)}$ with $j \neq i$,
normalized
such that ${\bf v}_{(i)}\cdot{\bf e}_{(j)}=\delta_{ij}$,
and let $\lambda_{(i)}$ be the eigenvalue associated with ${\bf e}_{(i)}$.
Using \mbox{\boldmath $\eta$}$^{(t)} = \sum_{i}
{\bf e}_{(i)}({\bf v}_{(i)}\cdot$\mbox{\boldmath $\eta$}$^{(t)})$
and noting that the eigenvalues and eigenvectors of ${\bf M}$ may be complex,
the left hand side of \eqref{egammadef} can be evaluated directly:
\begin{eqnarray}
\left\langle\frac{1}{L}\left|{\bf x}^{(T)}\right|^2\right\rangle 
   & = & \frac{1}{L}\left\langle\left|\sum_{t=0}^T 
                          {\bf M}^t\mbox{\boldmath $\eta$}^{(T-t)}\right|^2
                    \right\rangle  \\
\  & = & \frac{1}{L}\sum_{t=0}^T \sum_{s=0}^T \sum_{i=1}^L \sum_{j=1}^L 
                    (\lambda_{(i)})^t (\lambda_{(j)}^{\ast})^s 
                    \left({\bf e}_{(i)}\cdot{\bf e}_{(j)}^{\ast}\right) 
                    \left\langle ({\bf v}_{(i)}\cdot
			\mbox{\boldmath$\eta$}^{(T-t)})
                                 ({\bf v}_{(j)}^{\ast}\cdot
				\mbox{\boldmath$\eta$}^{(T-s)})
                    \right\rangle.
\end{eqnarray}
Using Eqs.~(\ref{eeta1}) and~(\ref{eeta2}) to evaluate the ensemble average
and performing the resulting geometric sum, we find
\begin{equation}
\lim_{T \to \infty} 
 \left\langle\frac{1}{L}\left|{\bf x}^{(T)}\right|^2\right\rangle 
    =  \frac{\sigma^2}{L}\sum_{i=1}^L \sum_{j=1}^L 
         \frac{1}{1-\lambda_{(i)}\lambda_{(j)}^{\ast}}
         \left({\bf e}_{(i)}\cdot{\bf e}_{(j)}^{\ast}\right)
         \left({\bf v}_{(i)}\cdot{\bf v}_{(j)}^{\ast}\right).
\end{equation}
Comparing to \eqref{egammadef} we find
\begin{equation} \label{egamma}
\gamma = \left(\frac{1}{L}\sum_{i=1}^L \sum_{j=1}^L 
         \frac{1}{1-\lambda_{(i)}\lambda_{(j)}^{\ast}}
         \left({\bf e}_{(i)}\cdot{\bf e}_{(j)}^{\ast}\right)
         \left({\bf v}_{(i)}\cdot{\bf v}_{(j)}^{\ast}\right)
         \right)^{1/2}.
\end{equation}

The added noise of strength $\sigma$ produces
fluctuations of magnitude $\gamma \sigma$ in the controlled
linear system.
Analytical estimation of $\gamma$ can be quite difficult, but exact
numerical evaluation of $\gamma$ is straightforward, given an explicit form
for ${\bf M}$.

Table~\ref{tgamma} shows values of $\gamma$ computed for various 
$L$, $\mu$, and $\epsilon$,
with ${\bf K}$ determined as described below.
The table also shows the maximum magnitude eigenvalue 
$\lambda_{(1)}$ for each case,
making it clear that
$\gamma$ may be large
even when all the $\lambda_{(i)}$'s are substantially less than unity.
The largeness of $\gamma$ derives from the nonnormality of
the ${\bf e}_{(i)}$'s,
which results in large magnitudes of some of the ${\bf v}_{(i)}$'s.
Again, the amplification of the noise is a purely linear effect directly
attributable to the transient growth of initial perturbations in nonnormal
systems.

It is important to note (though it may be obvious to some) that control
never fails in the purely linear system with noise added.
There can be no threshold above which the noise causes
divergence, since in the purely linear system there is no scale that
can determine such a threshold. 
Though the noise may be amplified substantially, it is always limited.

\section{Estimates of Tolerable Noise Levels}
\label{secestimate}
Given our exact computation of $\gamma$, we can now estimate
the value of $\sigma$ above which control will be lost in the
full nonlinear system.
Noting that the control perturbations are designed for
optimal stabilization of the linearized system,
we are led to consider the effect of the deviations from the linear behavior
due to nonlinear terms in the full equations.
We make the ansatz that correlations in these nonlinear deviations
may be neglected, and hence treat the nonlinear deviations as an
additional source of noise in the linear system.
We refer to the original noise of strength $\sigma$ as the ``additive noise''
and the deviations induced by nonlinearities as the ``deviational noise'' with
strength $\sigma_d$.

The size of the fluctuations about the fixed point will be given
approximately by 
\begin{equation}
\Delta = \gamma\sqrt{\sigma^2 + \sigma_d^2}.
\end{equation}
But $\sigma_d$ itself is produced by the nonlinear terms generated by the
fluctuations.
For the logistic map, $\sigma_d$ is therefore of the order of $\mu\Delta^2$,
so we have
\begin{equation}
\Delta = \gamma\sqrt{\sigma^2 + \mu^2\Delta^4}.
\end{equation}
This equation has a real solution for $\Delta$ if and only if
\begin{equation}\label{eestimate}
\sigma < \frac{1}{2\mu\gamma^2}.
\end{equation}
For $\sigma$ larger than this bound, the deviational noise
will exceed the additive noise, thereby generating even larger
deviational noise and an exponential divergence in the size of the
fluctuations.
Thus we take \eqref{eestimate} as our criterion for obtaining
effective control.

We note three reasons that this estimate could fail in principle.
First, the estimate assumes that the dominant nonlinearity encountered by
the fluctuating system is quadratic (which is true in the system studied
in this paper). 
If higher order nonlinearities become important before our criterion is
saturated, control may be lost for smaller $\sigma$.
Second, the estimate assumes that there are no correlations in
the deviational noise, which is not strictly correct.
Finally, it is possible that control would be lost for smaller $\sigma$
if the fluctuations are not distributed roughly evenly over the components 
of ${\bf x}$.
If a single component dominates the sum in \eqref{egamma}, for example,
it would be inappropriate to divide by $L$ in determining the relevant
size of the fluctuations.

For these reasons, it is necessary to investigate the accuracy of our estimate
using numerical simulation of the full nonlinear system with additive noise.
We have performed simulations on systems of size $L=10$ and $L=20$
for several different values of the parameters $\mu$ and $\epsilon$.
For each set of parameters values,
the feedback control matrix ${\bf K}$ was determined using standard
methods of linear-quadratic control theory. \cite{ogata1,ogata2} 
With weight matrices defined as
${\bf Q} = {\bf I}_{L \times L}$ 
and ${\bf R} = {\bf I}_{2 \times 2}$,
${\bf K}$ is obtained
from the relation:
\begin{equation}
{\bf K} = ({\bf R} + {\bf B}^\dagger {\bf PB})^{-1} {\bf B}^\dagger {\bf PA},
\end{equation}
where ${\bf P}$ is determined from the Ricatti equation:
\begin{equation}
{\bf P} = ({\bf Q} + {\bf A}^\dagger {\bf PA}) 
        - {\bf A}^\dagger {\bf PB}({\bf R} + {\bf B}^\dagger {\bf PB})^{-1}
        {\bf B}^\dagger {\bf PA},
\label{ricattieq}
\end{equation}
\eqref{ricattieq} is solved using a simple iterative procedure
until ${\bf P}$ converges according to the condition:
\begin{equation}
\frac{
\sum_{i,j} | P^{(T+1)}_{i,j} - P^{(T)}_{i,j} | }
{\sum_{i,j} |P^{(T+1)}_{i,j}|}
< 10^{-6}.
\end{equation}
More stringent convergence conditions did not produce noticeable changes
in ${\bf K}$.

Each run is started from the homogeneous
initial condition $z_i = z^\ast$ for all
sites $i$, and the full system with control is iterated 20,000 times.
All computations
(including the calculation of ${\bf K}$) are performed at a precision
of 30 decimal digits.  If control was lost during a run, the feedback
mechanism quickly caused the values of $z_i$ to stray from 
the allowed range for the logistic map and resulted in
the rapid divergence of $|z_i|$.
Figs.~\ref{fig3.1}--\ref{fig3.5}
shows $\sigma_{\rm max}$, the maximal value of the
noise strength that is effectively controlled, 
as a function of the coupling $\epsilon$ for
3 values of $\mu$.  
Our predictions of $\sigma_{\rm max}$ match
the measured values well for each $\mu$,
indicating that the criterion proposed in \eqref{eestimate}
captures the important physics of noise-induced loss of control.

\section{Comparison to method of Grigoriev, {\sl et al.}}

Grigoriev {\it et al.} have studied this problem from a different perspective.
They have developed an estimate for the noise level at which 
control fails based on an analysis
of the controllability of the linear system supplemented by
the assumption that the feedback signal applied cannot exceed 
a number of order unity \cite{grigorievprl}.

Briefly put, Grigoriev {\sl et al.}'s estimate relies on the 
well-known result that
controllability of a {\em noiseless} linear system implies the possibility
of directing the system from any arbitrary point in phase space
to the desired fixed point within a number of steps equal to the number 
of degrees of freedom $L$ \cite{ogata1,ogata2}.
This strict criterion is then adapted to the noisy case by
assuming that the relevant points in phase space for which
it must be possible to direct the system to the fixed point in
$L$ steps are just those that can be generated by iterating
the uncontrolled noisy system through $L$ steps.
If no constraints are placed on the size of the control perturbations,
the fact that the system is known to be controllable 
implies that this can always be done for the purely linear system,
regardless of the strength of the noise.
If the strength of the control perturbations are limited, however,
return to the fixed point in $L$ steps will not be possible for
sufficiently large noise strength.

In order to estimate the size of the control perturbations needed for
the particular geometry of the system at hand,
Grigoriev {\sl et al.} make the plausible assumption that every perturbation
which can affect the central site by the end of the $L$ iterations
should be of the order of magnitude required to produce an effect
of size $\lambda_{\rm max}^L\sigma$ after propagating to the central site,
where $\lambda_{\rm max}$ is the largest eigenvalue of ${\bf A}$.
The last perturbation that can propagate to the central site
by the end of the $L$ steps occurs at step $L/2$ and is 
suppressed (or amplified) by a factor $(\alpha\epsilon )^{L/2}$
before reaching the central site.
Letting the size of the perturbation be $u$,
we then have a necessary condition for effective control:
we must permit $u$ large enough such that 
\begin{equation}
u(\alpha\epsilon )^{L/2} = \lambda_{\rm max}^L\sigma.
\end{equation}
The estimate of the maximal $\sigma$ that can be controlled is
then given by the criterion that both $u$ and $u(\alpha\epsilon )^{L/2}$
remain less than unity.
This argument leads directly to the criteria published
in Ref.~\cite{grigorievprl}.
Note that for the system of coupled logistic maps, 
it is not clear from this analysis
whether the restriction that  control perturbations 
must not exceed order unity arises due to the fact that
the individual maps diverge rapidly for $x$ outside the unit interval
or due to the fact that nonlinear effects become important.
The  analysis presented in Section~\ref{secestimate} above
clearly indicates that it is the latter effect that is most important.

The method based on controllability has one great advantage over ours 
in that it makes no reference to the matrix ${\bf K}$;
it is intended to apply to the optimal choice of ${\bf K}$,
which may be different from the one determined above.
One cannot rule out, for example, the possibility that a
different choice of ${\bf K}$ would permit control of significantly
higher noise levels.

\begin{figure}
\centerline{\epsfbox{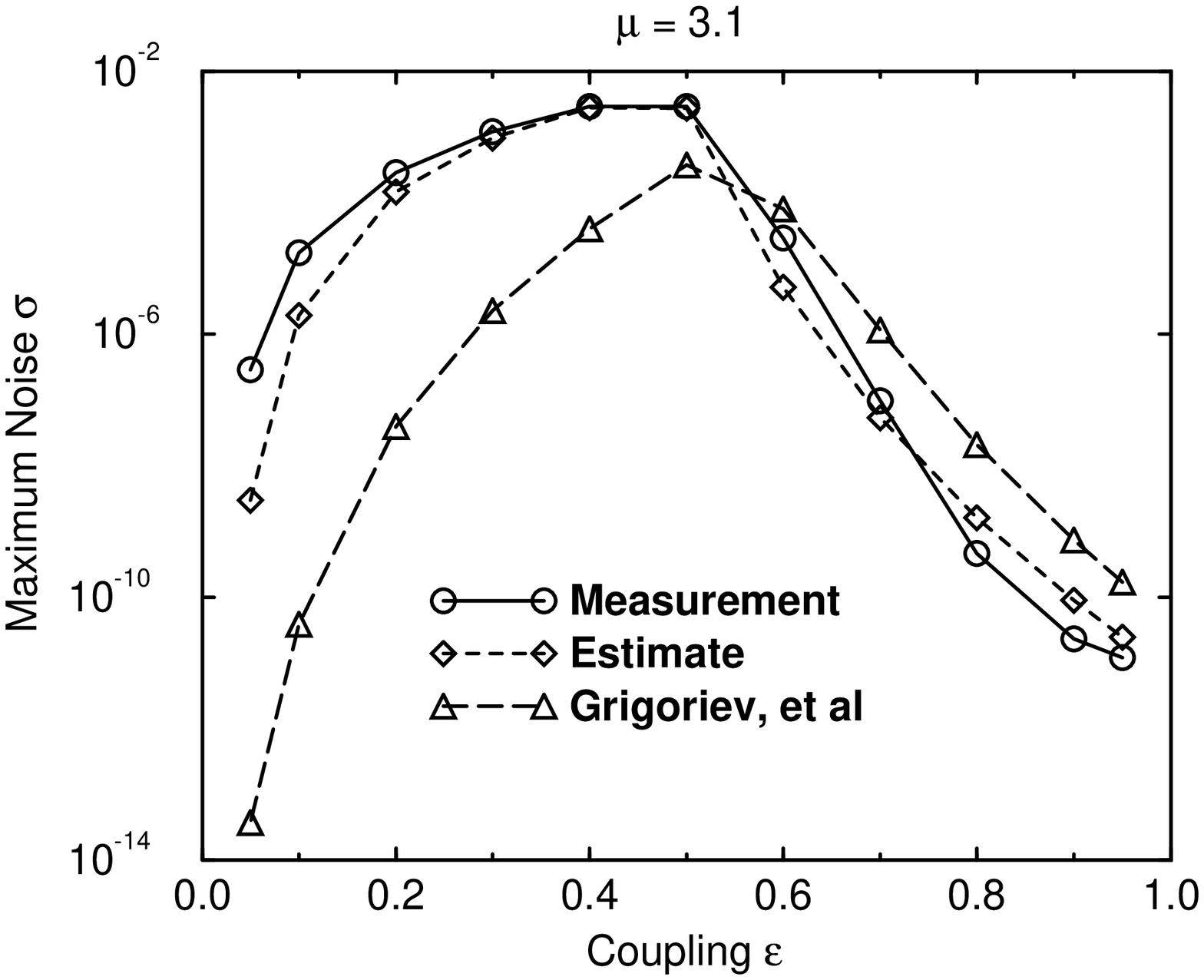}}
\caption{The measured value, the prediction of \eqref{eestimate}, and
the prediction of Grigoriev, {\sl et al.} for the maximal noise
strength $\sigma_{\rm max}$
for which control can be achieved for $\mu = 3.1$ and $L = 20$.
}
\label{fig3.1}
\end{figure}

\begin{figure}
\centerline{\epsfbox{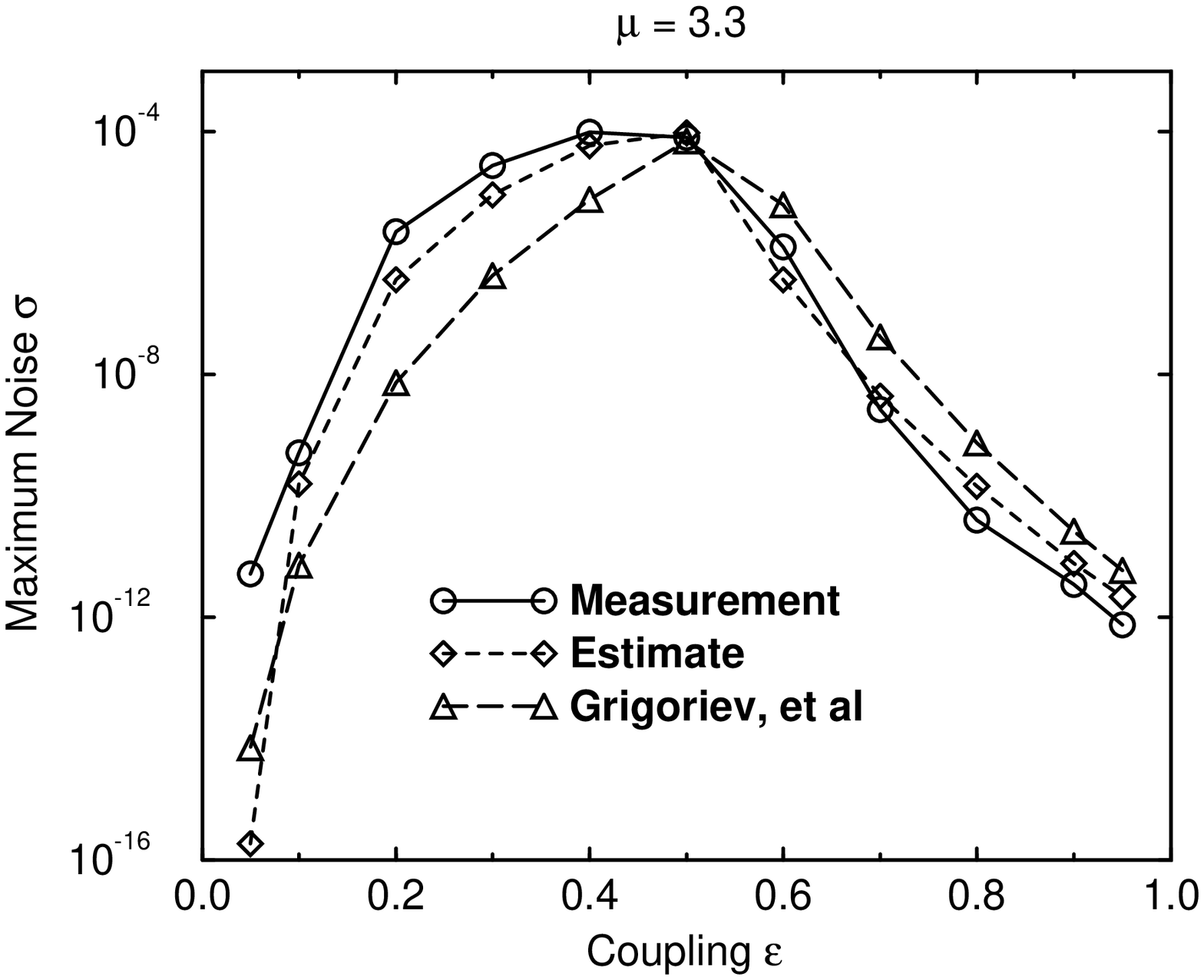}}
\caption{The measured value, the prediction of \eqref{eestimate}, and
the prediction of Grigoriev, {\sl et al.} for the maximal noise
strength $\sigma_{\rm max}$
for which control can be achieved for $\mu = 3.3$ and $L = 20$.
}
\label{fig3.3}
\end{figure}

\begin{figure}
\centerline{\epsfbox{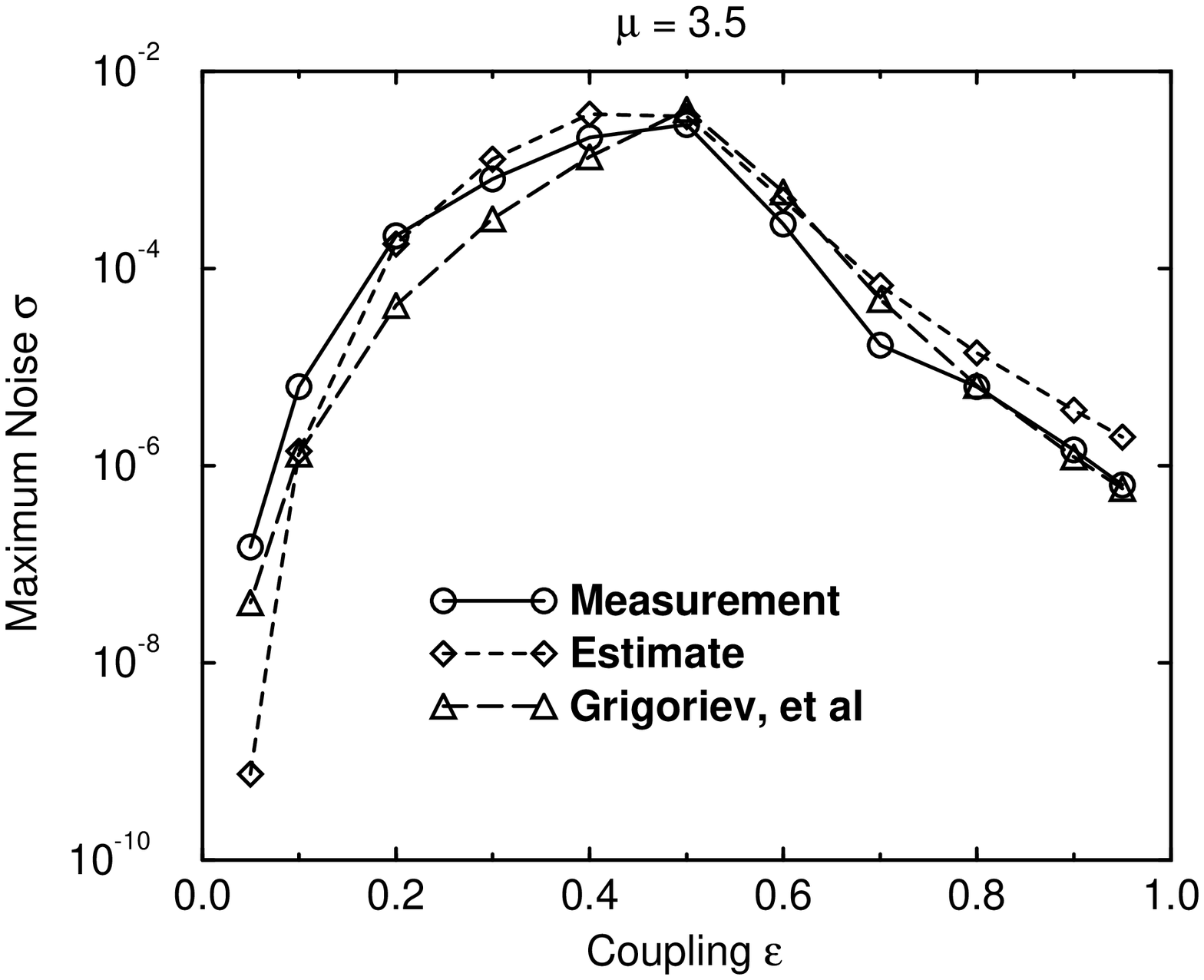}}
\caption{The measured value, the prediction of \eqref{eestimate}, and
the prediction of Grigoriev, {\sl et al.} for the maximal noise
strength $\sigma_{\rm max}$
for which control can be achieved for $\mu = 3.5$ and $L = 10$.
}
\label{fig3.5}
\end{figure}

The price of the generality of the method is inaccuracy in certain classes
of systems.
As shown in Figs.~\ref{fig3.1}--\ref{fig3.5}, 
for example, 
the prediction of Grigoriev {\sl et al.} 
substantially underestimates the noise level that can
be tolerated for small values of the coupling.  
For $\epsilon > 0.5$, however, the prediction of Grigoriev, {\sl et al.}
tends to overestimate the maximum controllable noise strength.

The estimate of 
Grigoriev {\sl et al.} handles the
linear aspects of the problem quite elegantly but uses a rather
crude estimate of when the nonlinear effects become important.
It is possible for the deviational noise due to the nonlinearity
to become important for much lower additive noise strengths than those
required to force control perturbations of order unity.
By directly computing the amplification of the noise by a
specific proposed controller,
we arrive at a more accurate estimate of the point at which
nonlinear effects will become important and thereby invalidate
the linear analysis.
Note that the loss of control has nothing to do
with the inability of the controller to supply sufficiently large
perturbations, as might be suggested by the use of a cutoff of
unity for the feedback signals in the analysis of
Grigoriev, {\sl et al}.
Nor is it correct to estimate the importance of the nonlinear effects
simply by comparing the magnitudes of linear and nonlinear terms
on a single iteration.
Rather, the loss of control is due to the effect described
in Section~\ref{secestimate}
above, and occurs for feedback levels much lower than unity.

We view our results as complementary to those of Grigoriev {\sl et al.},
both for practical and conceptual reasons.
Taken together, they form a coherent picture of the breakdown
of control in a spatially extended homogeneous system.
The important conceptual points can be summarized as follows:
(1) sparsely distributed controllers in such systems give rise to 
highly nonnormal eigenvectors in the vicinity of a homogeneous fixed point;
(2) this induces a large amplification of both the noise and the 
nonlinear deviations from the linearized systems; and
(3) an accurate estimate of the tolerable noise level for a given 
implementation of feedback control can be obtained by considering
the nonlinear effects as an additional source of noise and applying
the criterion of Eq.~(\ref{eestimate}).

\acknowledgements{We thank R. Grigoriev and J. Doyle for useful conversations.
This work was supported by the National Science Foundation under
grants DMR-9705410, DMR-9419506, and DMR-9412416.}


\begin{thebibliography}{1}

\bibitem{huprl}
G. Hu and Z. Qu, Phys. Rev. Lett. {\bf 72},  68  (1994).

\bibitem{grigorievprl}
R.~O. Grigoriev, M.~C. Cross, and H.~G. Schuster, Phys. Rev. Lett. {\bf 79},
  2795  (1997).

\bibitem{doyle}
J.~C. Doyle, B.~A. Francis, and A.~R. Tannenbaum, {\em Feedback control theory}
  (Macmillen Publishing Co., New York, 1992).

\bibitem{ogata1}
K. Ogata, {\em Discrete-time control systems} (Prentice-Hall, Englewood Cliffs,
  NJ, 1995).

\bibitem{ogata2}
K. Ogata, {\em Modern Control Engineering} (Prentice-Hall, Englewood Cliffs,
  NJ, 1997).

\bibitem{trefethen1}
L.~N. Trefethen, A.~E. Trefethen, S.~C. Reddy, and T.~A. Driscoll, Science {\bf
  261},  578  (1993).

\bibitem{trefethen3}
K.-C. Toh and L. Trefethen, SIAM J. Sci. Computing {\bf 17},  1  (1996).

\end{thebibliography}

\newpage

\begin{table}
\begin{tabular}{|c|c|c|c|c|} \hline
$\mu$ & $\epsilon$ & $L$ & $\gamma$ & $|\lambda_{(1)}|$ \\ \hline
3.5 & 0.30 & 10 & 1122. & 0.807 \\ \hline
3.5 & 0.50 & 10 & 412. & 0.796 \\ \hline
3.5 & 0.70 & 10 & 21427. & 0.826 \\ \hline
3.3 & 0.30 & 20 & 328814. & 0.915 \\ \hline
3.3 & 0.50 & 20 & 31615. & 0.906 \\ \hline
3.3 & 0.70 & 20 & 6.79 $\times 10^8$ & 0.886 \\ \hline
3.1 & 0.30 & 20 & 3362. & 0.953 \\ \hline
3.1 & 0.50 & 20 & 1174. & 0.945 \\ \hline
3.1 & 0.50 & 20 & 5.98 $\times 10^7$ & 0.946 \\ \hline
\end{tabular}
\caption{The amplification factor $\gamma$ and the maximum
magnitude eigenvalue $\lambda_{(1)}$ for a representative set of
parameters $\mu$, $\epsilon$, and $L$.}
\label{tgamma}
\end{table}

\newpage

\end{document}